\documentclass[11pt]{article}
\newcommand{\be}{\begin{eqnarray}}\newcommand{\beq}{\begin{equation}}
\newcommand{\ee}{\end{eqnarray}}\newcommand{\eeq}{\end{equation}}
\newcommand{\De}{\Delta}
\newcommand{\la}{\lambda}
\input epsf 
\input amssym.def 
\input amssym.tex 
\usepackage{graphicx} 
\title{ %A model for the nucleation mechanism of protein folding 
Role of the dihedral angle potential in the nucleation pathway of protein folding
}
\author{Y. S. Djikaev\thanks{E-mail: idjikaev@eng.buffalo.edu}
\hspace{0.2cm}  
\\ Department of Chemical and Biological  Engineering, SUNY at Buffalo, \\ 
Buffalo, New York  14260 }

\date{(Received\hfill .}
\renewcommand{\baselinestretch}{1}
\begin{document}
\renewcommand{\baselinestretch}{1}
\maketitle
\renewcommand{\baselinestretch}{1}
{\bf Abstract.} 
{\small 

A kinetic model for the nucleation mechanism of protein folding is proposed. A 
protein is modeled as a heteropolymer consisting of hydrophobic and hydrophilic
beads with equal  constant bond lengths and bond angles. The total energy of
the heteropolymer is determined by the repulsive/attractive  interactions of
non-linked beads and the contribution from the dihedral angles involved. Their 
parameters can be rigorously defined, unlike the  ill defined surface tension
of a cluster of protein residues which is the basis of the previous  model. As
a crucial idea of the model, the dihedral potential in which a selected bead is
involved is averaged over all  possible configurations of neighboring beads
along the protein chain. The resulting average dihedral potential of the
residue is constant far enough from the cluster, but increases monotonically
with decreasing distance below a threshold value. An overall potential around
the cluster wherein a residue performs a chaotic motion is a combination of the
average dihedral and pairwise potentials. As a function of distance from the
cluster it has a double well shape. Residues in the inner  well are considered
as belonging to the cluster (folded part of the protein) while those in the
outer well are treated as belonging to the unfolded (although compact) part of
the protein. A double well shape of the potential around the cluster allows one
to determine its emission and absorption rates by using a first passage time
analysis and develop a  self-consistent kinetic theory for the nucleation
mechanism of protein folding. Numerical calculations for a protein of 2500
residues with the diffusion coefficient of residues in the native state ranging
from $10^{-6}$ cm$^2$/s to $10^{-8}$ cm$^2$/s predict folding times in the range from
several seconds to several hundreds of seconds.

} 

\renewcommand{\baselinestretch}{1} 
\newpage 
\section{Introduction}
\renewcommand{\baselinestretch}{1}

\par Proteins play an overwhelmingly dominant role in life. If a specific job has to by done in a
living organism, it is almost always a protein that does it. Life depends on thousands of different
proteins whose structures are fashioned so that individual protein molecules combine, with exquisite
precision, with other molecules. In order for a protein molecule to carry out a specific biological
function, it has to adopt a well-defined three-dimensional structure.$^{1,2}$ The formation of this
structure (of a biologically active globular protein) constitutes the core of 
a so-called ``protein folding problem".$^{3}$ Many thermodynamic and kinetic aspects of the process
remain obscure and its mechanism elusive.$^{4-8}$

\par Experiment and simulation suggest that there exist multiple pathways for the protein 
folding.$^{6-15}$ It is believed that initially a denatured protein very quickly transforms into a 
compact (but not native) configuration with a few, insignificant amount of tertiary contacts.  The
transition from such a compact configuration to the native one has been suggested  to occur via  two
distinct mechanisms.  One of them can be referred to as a ``transition state mechanism" whereby the 
tertiary contacts of the native structure form as the protein passes through a sequence of
intermediate states thus  gradually achieving its unique spatial configuration.$^{6-15}$ The protein
in intermediate states has a native-like  overall topology but is stabilized by incorrect hydrophobic
contacts. These states correspond to  misfolded forms of the native protein. The transition from
these intermediate,  misfolded states to the correctly folded,  native structure is a slow process
(because it involves a large-scale rearrangement of the molecule) occurring on a relatively large
time scales.$^{14,15}$  Alternatively, the transition  from the compact ``amorphous" configuration
to the native state occurs immediately following the formation of some number of tertiary
contacts.$^{14,15}$  This mechanism is similar to nucleation, i.e., once a critical number of
(native) tertiary contacts is established the native structure is formed without passing through any
detectable intermediate states.$^{14}$ 

So far most of the work on protein folding has been done by using either Monte Carlo (MC) or molecular
dynamics (MD) simulations. A rigorous theoretical  treatment of the problem by means of the statistical
mechanics is hardly practicable because of the extreme complexity of the system, although some
approximate treatments were already  reported.$^{16,17}$ A theoretical model for the nucleation
mechanism of the process has so far remained underdeveloped.$^{18,14,19}$  The model which existed so
far is a thermodynamic one considering the formation of a cluster of protein residues and  calculating
its free  energy change, much like the classical nucleation theory (CNT) does. The cluster  is
characterized by $\nu$, the total number of residues, with the mole fraction of hydrophobic ones assumed
to be known. As usual in CNT, the size of a critical cluster (nucleus) is provided by the location of
the maximum of  the free energy of formation as a function of $\nu$. Such an approach necessarily
involves  the concept of surface tension for a cluster consisting of protein residues.  (Clearly, this
quantity is an intrinsically ill-defined physical quantity and can be considerer only as  an adjustable
parameter; no direct experimental measurement thereof is possible.)  After the formation of the nucleus
(critical size cluster of residues),  the protein quickly reaches its native state. 

In what follows work I present a new, microscopic model for the nucleation mechanism of the protein
folding. The new model is based on ``molecular" interactions, both long-range (due to 
repulsion/attraction) and configurational (due bond and dihedral angles), in which protein residues are
involved. Their parameters can be rigorously defined, and it should be possible (although not
straightforward) to determine them theoretically, computationally, or experimentally.  The ill-defined
surface tension of a cluster of protein residues does not enter into  the new model which is thus  more
advanced than the CNT-based one. The crucial idea underlying the new model consists of averaging the
dihedral potential in which a selected residue is involved over all the possible configurations of
neighboring residues. The resulting average dihedral potential depends on the distance between residue
and cluster. Its combination with the average long range  potential (due to pairwise interactions of the
selected residue with those in the cluster) gives rise to the overall potential which has generates a
pair of potential wells around the cluster  with a barrier between them. Residues in the inner well are
considered to belong to the cluster (part of the protein with correct tertiary contacts)  while those in
the outer well are treated as belonging to the mother phase (amorphous  part of the protein with
incorrect tertiary contacts). Transitions of residues from the inner well into the outer one and vice
versa are considered as elementary emission and absorption events, respectively. The rates of emission
and absorption of residues by the cluster are determined by using the first passage time
analysis.$^{20-25}$  Once these rates are found as functions of the cluster size, one can develop a
self-consistent kinetic  theory for the nucleation mechanism of folding of a protein. For example,  the
size of the critical cluster (nucleus) is then found as  the one for which these rates are equal. The
time necessary for the protein to fold can be evaluated as a sum of the times necessary for the
appearance of the first nucleus and the time necessary for the nucleus to grow to the maximum size (of
the folded protein in the native state). 

The paper is structured as follows. In Section 2 we describe a random heteropolymer chain whereby  a
protein molecule is often modeled$^{6,14}$ and outline a CNT-based model for the nucleation mechanism
of protein folding, whereof a new, microscopic model is proposed in Section 3. The results of
numerical calculations are presented in Section 4 and a brief discussion and conclusions are summarized
in Section 5.

\section{Heteroplymer chain as a protein model and a CNT based model for the nucleation mechanism of 
protein folding}

\subsection{A heteropolymer as a protein model}

As a simple model of a protein in MD and MC simulations  of
protein folding dynamics, the polypeptide chain of a protein is considered$^{6,14}$ as a heteropolymer
consisting of $N$ connected beads which can be thought of as representing the $\alpha$-carbons of
various amino acids. The heteropolymer may consist of hydrophobic ($b$), hydrophilic ($l$), or neutral
($n$).  Two adjacent beads are connected by a covalent bond of fixed length $\eta$. This model (and its
variants),  augmented with appropriate interaction, bond, and dihedral potentials described below, was
shown$^{6,14,16,18}$ to be capable of capturing the essential characteristics of protein folding
process even though it contains only some features of a real polypeptide chain. For example, this model
ignores side groups although they are known to be crucial for intramolecular hydrogen bonding.$^{1}$
Besides, the presence of solvent (water) in a real physical system has been 
usually accounted for too simplistically, although the protein dynamics was reported to be more
realistic in MD simulations where solvent molecules are explicitly
present.$^{26}$ Despite these limitations, various modifications of 
heteropolymer models$^{5,16,18,27-30}$ shed light on some 
important details of the folding of polypeptide chains, such as possible pathways for the protein
transition from a denatured state to the native one.$^{6,14}$

The total energy of the heteropolymer (polypeptide chain) can contain the contributions of three
different types. First, the contribution from repulsive/attractive forces between pairs of non-adjacent
beads (these can be, e.g., of Lennard-Jones type or others). The next  contribution can arise from
harmonic forces related with the oscillations of  bond angles. Finally, there is a contribution
from the dihedral angle potential due to the rotation around the peptide bonds. There are various ways
to model these three types of energetic terms.$^{6,14,27-29}$  

A pair interaction between two non-adjacent beads $i$ and $j$ at a distance $r_{ij}$ away from each other 
can be taken, for example, as$^{6,14}$ 
\beq \phi_{ij}(r_{ij})=\left\{
\begin{array}{ll} 
4\epsilon_b [(\eta/r_{ij})^{12}-(\eta/r_{ij})^{6}] & (i,j=b),\\
4\epsilon_l [(\eta/r_{ij})^{12}+(\eta/r_{ij})^{6}] & (i=l,j=b,l),\\
4\epsilon_n (\eta/r_{ij})^{12} & (i=n,j=b,l,n),\\
\end{array}
\right. \eeq
where $\eta$ is the bond length (fixed) and $\epsilon_b,\epsilon_l,$ and $\epsilon_n$ are the energy
parameters.

The angle $\beta$ between two successive bonds (in the heteropolymer) can be regarded to 
be subjected to a harmonic potential  
\beq \phi_{\beta}=\frac{k_{\beta}}{2}(\beta-\beta_0),\eeq
where the spring constant $k_{\beta}$  is relatively large 
(in refs.6,14 it was taken $20\epsilon_b$/(rad)$^2=105^{\small o}$
so that the deviation of the bond angles from the average value $\beta_0$ is very small.
Hence  all bond angles can be set to be equal to $\beta_0$ (as argued in refs.6,14, the bond angle
forces play a minor role in the protein folding/unfolding). 

The dihedral angle potential arises due to the rotation of three successive peptide bonds connecting
four successive beads, and is related to the dihedral angle $\delta$ as 
\beq \phi_{\delta}=\epsilon'_{\delta}(1+\cos{\delta})+\epsilon''_{\delta}(1+\cos{3\delta}),\eeq 
where $\epsilon'_{\delta}$ and $\epsilon''_{\delta}$  are independent energy parameters.
This potential has three minima, one in the {\em trans} configuration at $\delta=0$ and two others
in the {\em gauche} configurations at $\delta=\pm\arccos{\sqrt{(3
\epsilon''_{\delta}-\epsilon'_{\delta})/12\epsilon''_{\delta}}}$ (the former one 
is lower than the latter two).

The above structure of potential functions for a heteropolymer was suggested 
by Honeycutt and Thirumalai,$^{6}$ 
while Bryngelson and Wolynes$^{16,18}$ used a random energy model and 
Skolnik and co-workers$^{27-29}$ developed 
discrete analogs (for a diamond lattice) of eqs.(1) and (3)
augmented  with a ``cooperativity potential" as a crucial element of the model.
It was shown$^{6,14}$ by MD simulations (employing low friction Langevin dynamics) that a proper balance
between the above three contributions to the total energy of the heteropolymer ensures that the
heteropolymer folds into a well defined $\beta$-barrel structure. The balancing between these terms
is performed by adjusting the energy parameters
$\epsilon_b,\epsilon_l,\epsilon_n,\epsilon'_{\delta},\epsilon'_{\delta}$ for each type of beads. It
was also found$^{6.14}$ that the  balance between the dihedral angle potential, which tends to
stretch the molecule into a state with all bonds in a {\em trans} configuration, and the attractive
hydrophobic potential is crucial to induce folding int a $\beta$-barrel like structure upon
cooling. Excessively dominant attractive forces make the heteropolymer fold into a globule-like
structure, while an overwhelming dihedral angle potential makes the chain remain in an unfolded
(elongated) state (even at low temperatures) with bonds mainly in the {\em trans} configuration.

A possibility that the nucleation-like mechanism can constitute the most viable pathway for the
protein folding was first suggested by Guo and Thirumalai$^{14}$ 
(although Bringelson and Wolynes$^{18}$ also drew 
the analogy  between the results presented therein and the thermodynamics of cluster
formation in the framework of CNT). The formalism of the nucleation mechanism for the protein folding is
invoked to evaluate the size of the critical cluster (nucleus) of native protein residues (whereof the
formation leads to a quick transition of the whole protein into its native state). Denoting the total
number of residues in the protein by $N_0$, let us consider a formation of a cluster having a correct
tertiary structure in an unfolded protein. The free energy of formation of such a cluster of $\nu$ native
residues (i.e., residues which are in the same state as they are in the native protein) can be written (in
the framework of CNT) as 
\beq W = -\nu\De\mu  + \sigma 4\pi\lambda^2\nu^{2/3},\eeq
where $\delta\mu\equiv\mu_d-\mu_n$ is the difference between the free energy per residue in the
denatured and native states,respectively (marked with the subscripts ``d" and ``n"), $\sigma$ is
the ``surface"  tension (energy) of the boundary between the cluster (having a native structure)
and the  unfolded part of the protein, $\lambda=\left(3v/4\pi\right)^{1/3}$, and $v$ is a volume
per protein residue in its native state. 

In ref.14 it was argued that the initial stage of the protein folding is driven by a hydrophobic
attractive forces so that the volume term (i.e., the first one) in eq.(4) was determined by the
number of hydrophobic contacts in the cluster and hence could be specified as $-(1/2)\epsilon_b
\chi \nu(\chi\nu-1)$, where $\chi$ is the mole fraction of hydrophobic residues in the cluster
(assumed the same as in the whole protein). As a result the number of residues in the critical
cluster was given as  $\nu_c=(8\pi\sigma \lambda^2/3\chi^2\epsilon_b)^(3/4)$ which for typical
values of $\lambda, \sigma$, and $\epsilon_b$ was estimated$^{14}$ to be of the order of $10$. In
ref.16, $\Delta\mu$ in the volume term of eq.(4) was evaluated to be of the order of $0.1k_BT$
($k_B$ is the Boltzmann constant, and $T$ is the temperature). The ``surface" tension was argued to
arise because the amino acid residues located at the cluster surface interact stronger with the
cluster interior than with the unfolded part of the protein. Since the interaction energies in
protein folding are of the order of $k_BT$, the surface tension $\sigma$ multiplied by $4\pi\la^2$ was estimated to be of the
same order and the number of residues in the critical cluster was evaluated$^{18}$ to be of the
order of $100$ (for $N_0=150$). Both estimates 
corroborate the idea that the nucleation mechanism can constitute a viable pathway for the protein
folding.$^{31-35}$ 

\section{Kinetics of nucleation during protein folding}

The CNT-based model for the nucleation mechanism of protein folding is limited to its  thermodynamics, namely to the free energy of formation of the cluster of native residues. For a system in  the thermodynamic limit (both the number of molecules
$N\rightarrow \infty$  and the volume $V\rightarrow\infty$), the validity of expression (4) for the
free energy of cluster formation in various [i.e., canonical (NVT), grand canonical ($\mu$VT), and
Gibbs (NPT)] ensembles was well established.$^{36,37}$ If nucleation occurs in a finite  size system, there appear 
additional terms on the RHS of eq.(4) which depend not only on the size of the system but also on
the nature of the ensemble.  However, a folding protein (mostly containing much less
than a couple of thousands of amino acids)  can hardly be considered to satisfy the thermodynamic
limit. Furthermore,  the cluster formation during protein folding occurs under conditions which
cannot be identified with either of commonly used thermodynamic ensembles. Besides,  the CNT based
model (described above) has inherited a complicated problem of CNT  related to the surface tension
of the cluster.$^{}$  It was argued that the concept of surface tension   may not be  adequate for
too small clusters (such as those of interest in nucleation),$^{38,39}$  not to mention  the
assumption (of CNT)  that it is equal to the surface tension of a planar  interface. Although CNT
produces reasonable agreement with experiment on unary nucleation, its application to
multicomponent nucleation leads to several inconsistencies and large discrepancies with
experimental data$^{40-45}$ which are  blamed on the inadequate use of the concept of surface
tension. In the case of protein folding this problem is even more complicated because $\sigma$ in
eqs.(4) is an ill-defined  quantity which is  experimentally impossible to determine due to the
non-existence  of bulk ``folded protein" and ``unfolded protein" as real physical  phases, not to
mention a flat interface between them. 

\par In order to avoid the use of macroscopic thermodynamics in the kinetic theory of unary
nucleation,  an alternative approach was proposed$^{20-22}$ on the basis of the mean first passage
time  analysis. Unlike CNT, that theory$^{20-22}$  is built upon molecular interactions and
does not make use of the free energy of formation of  tiny clusters involved in nucleation. Instead,
the theory$^{20-22}$ exploits the fact that  one can derive and solve the kinetic equation of
nucleation (hence find the nucleation rate)  if the emission and absorption rates of a cluster are
known as functions of its size. For the rate of absorption of molecules by the cluster, the new
approach uses (as CNT does) a standard gas-kinetic expression,$^{46}$ but the rate of emission of molecules by the cluster is determined via a
mean first  passage time analysis. This time is calculated by solving a single-molecule  master
equation for the  probability distribution function of a surface layer molecule moving in a potential
well around the cluster.   The master equation is a Fokker-Planck equation in the phase space which
can be reduced to the Smoluchowski equation owing to the hierarchy of characteristic time scales in
the evolution of the single-molecule distribution function  with respect to coordinates and
momenta.$^{20-22}$ 
Recently, a further development of that kinetic theory was proposed by combining it 
with the density functional theory (DFT)$^{23,24}$ and extending it to binary$^{24}$ and heterogeneous$^{25}$  systems. 

Note that although the emission rate of the cluster in refs.20-25 was found by using a first passage
time analysis, for the absorption rate there was used an expression derived in the framework of  the
gas-kinetic theory of gases$^{46}$ which assumes a Maxwellian distribution of velocities of mother
phase molecules. This assumption being unquestionably valid for vapor-to-liquid nucleation in dilute
(if not ideal) gases,  becomes increasingly inaccurate as the density of the mother phase increases
and molecular  interactions therein become non-negligible. Clearly, this assumption (hence the
absorption rate based thereupon) is inadequate in considering the cluster formation during the
protein folding. Indeed, the amino acid residues of the protein are all successively linked by bonds
of virtually fixed length each and fixed angle between each pair. 

In this section we will present a new, kinetic model for the nucleation mechanism of protein folding
based on the first passage analysis which will be used for determining not only the rate of emission
(of native residues from the cluster) but also the rate of absorption (of
non-native residues by the cluster). The general formalism of our model is a mean  first passage
time  analysis, but a crucial modification (compared to refs.20-25) must be introduced thereto in
order to make it applicable to nucleation in a protein. This modification concerns the potential well
generated around the cluster  as a result of all its interactions with a residue which moves around
the cluster while being a part  of the protein backbone (a bead in a heteropolymer).

\subsection{Potential well around a cluster within a protein}

A heteropolymer chain as a protein model, originally proposed in refs.6,14 and described above, consists of three types of beads - neutral, hydrophobic, and hydrophilic. The neutral beads play an important role in that model. Their interaction with each other is purely repulsive and the dihedral angle 
forces are assumed to be weaker for the bonds involving them so that the bend formation is enhanced in
regions where they are present. 
In real proteins this kind of residues can be thought to ensure  the formation of loops and turns. 
MD simulations$^{6,14}$ show that such a heteropolymer acquires a $\beta$-barrel shape in the
lowest energy structure, with neutral residues appearing mostly in bend regions. This work is not aimed at obtaining a $\beta$-barrel structure of the folded protein, so neutral beads will be removed from the model. Clearly, this will
require to rebalance $\epsilon$'s in eqs.(1),(3) in order to facilitate the 
formation of loops and turns in a heteropolymer chain. 

Thus, the two-component heteropolymer chain  as a model for a protein consists of only hydrophobic 
and hydrophilic beads without neutral ones with the pair interaction, bond angle, and dihedral angle
potentials given by eqs.(1)-(3). With this 
assumption, the formation of a cluster consisting of native residues during the protein folding can
be regarded as binary nucleation. We shall therefore present a model for the nucleation mechanism of
protein folding in terms of binary nucleation by using a first passage time analysis$^{20-25}$ with 
a crucial modification concerning the potential well around the cluster. 

\par Consider a binary cluster of spherical shape (with sharp boundaries and radius $R$) 
immersed in a binary fluid mixture.$^{24}$ A molecule of component $i\;\;(i=b,l)$ located in the 
surface layer of the cluster was considered to perform thermal chaotic motion in a spherically symmetric
potential well $\phi_i(r)$ resulting from the pair interactions of this molecule with those in the
cluster. 
Assuming pairwise additivity of the intermolecular 
interactions, $\phi_i(r)$ is provided by 
\beq \phi_i(r)=\sum_{j}\int_V\;d{\bf r'}\;\rho_j(r')\phi_{ij}(|{\bf r'}-{\bf r}|),\eeq 
Here ${\bf r}$ is the coordinate of the surface molecule $i$, $\rho_j(r)\;\;(j=1,2)$  is the number density
of molecules of component $j$  at point ${\bf r'}$ (spherical symmetry is assumed, the cluster center
chosen as the origin of the coordinate system), and
$\phi_{ij}(|{\bf r'}-{\bf r}|)$ is the interaction potential between two molecules of components $i$
and $j$ at points ${\bf r}$ and ${\bf r'}$, respectively.  The integration in eq.(5) goes over the  whole 
volume of the system, but the vapor phase contribution can be assumed to be small and accounted for by a 
particular choice of the $\epsilon_b$ and $\epsilon_l$. 

\par For nucleation in proteins the potential $\psi_i(r)$ for a residue of type $i$ around the 
cluster is determined not only by the potential $\phi_i(r)$, but also by two other contributions, 
$\phi_{\beta}(r)$ and $\bar{\phi}_{\delta}(r)$, due to the bond angle and dihedral angle
potentials, respectively:
$$ \psi_i(r)=\phi_i(r)+\phi_{\beta}(r)+\bar{\phi}_{\delta}(r).$$ 
Without affecting the generality of the model, one can  significantly simplify the algebra and eventual
numerical calculations by  assuming that all bond angles are fixed and equal to $\beta_0=105^{\small
o}$. Under this assumption the contribution to the potential energy of the protein arising from the
bond  angle potential is constant and does not depend on the distance $r$ between the selected bead and
the center of the cluster. Therefore, the term $\phi_{\beta}(r)$ on the RHS of eq.(6) can be disregarded
(or, equivalently, be chosen as a reference level for the potential energy), i.e.,   
\beq \psi_i(r)=\phi_i(r)+\bar{\phi}_{\delta}(r).\eeq 

The term $\phi'_{\delta}(r)\equiv\phi'_{\delta}(r,{\bf r_2},{\bf r_3},{\bf
r_4},{\bf r_5},{\bf r_6},{\bf r_7})$ in $\psi_i(r)$ is due to the dihedral angle potential of the whole
protein. Consider bead $1$ (of type $b$ or $l$) at a distance $r$ from the center of the cluster (see
Figure 1). The total dihedral angle potential of the whole protein chain  for a given configuration of
beads 2,3,...,N can be written in the form 
\beq \phi'_{\delta}(r)=\phi_{\delta}(\delta^{642}_{421}(r))+\phi_{\delta}(\delta^{421}_{213}(r))+
\phi_{\delta}(\delta^{213}_{135}(r))+\phi_{\delta}(\delta^{135}_{357}(r)).\eeq
where $\delta^{ijk}_{jkl}$ is a dihedral angle between two planes, one of which is 
determined by beads $i,j,k$ and the other by beads $j,k,l$. 
On the RHS of the above equation, an independent of $r$ term is omitted which 
represents the contributions from the dihedral angles involving beads $8,9,...,N$ (hence
$\phi'_{\delta}(r)$ does not depend on coordinates of beads $8,...,N_0$). 
It can be regarded as affecting only the reference level for $\psi_i(r)$. 

Consider bead $1$ at a given distance from the cluster $r$. Various configurations of beads 
$2,3,...,N$ (subject to the fixed bond length and bond
angle constraints as well as to the constraint of excluded cluster volume) lead to various sets of
dihedral angles. However, variations in the location of beads $8,9,...,N$  lead to variations in the
dihedral potential which are independent of $r$.  Thus, the dihedral term $\bar{\phi}_{\delta}(r)$ (less an independent of $r$ term omitted hereafter) on the RHS of eq.(7)  can be obtained by 
averaging eq.(8) with the probability distribution function $p({\bf r},{\bf r_2},{\bf r_3},{\bf r_4},{\bf r_5},{\bf r_6},{\bf r_7})$ for configurations of beads $2$ to $7$ with a fixed location of bead $1$ and assigning the result to the latter: 
\beq \bar{\phi}_{\delta}(r)=
\int_{\Omega_{18}}d{\bf r_2}d{\bf r_3}d{\bf r_4}d{\bf r_5}d{\bf r_6}d{\bf r_7}\; 
\phi'_{\delta}(r)p({\bf r},{\bf r_2},{\bf r_3},{\bf r_4},{\bf r_5},{\bf r_6},{\bf r_7})\eeq 
where ${\bf r_i}\;\;(i=2,...,7)$ is the radius-vector of bead $i$ and $\Omega_{18}$ is the integration region in an 18-dimensional space. 

It is convenient to 
choose a Cartesian system of coordinates with the origin in the cluster center in such a way that
the coordinates of bead 1 are $x_1=0,y_1=0,z_1=r$ (see Figure 1). The Cartesian coordinates of other beads will be
denoted  by $x_i,y_i,z_i\;\;(i=2,...,7)$. The Cartesian coordinates of bead 2 are related to its
spherical  ones $r_2,\Theta_2,\varphi_2$ by 
\beq x_2=r_2\sin{\Theta_2}\cos{\varphi},\;\;y_2=r_2\sin{\Theta_2}\sin{\varphi},\;\;z_2=r_2\cos{\Theta_2}.
\eeq

At a given $r$ (the location of bead 1 is fixed), the polar angle $\Theta$ of bead 2 
is uniquely determined by $r_2$ due to the constant bond
length constraint: 
\beq \tilde{\Theta}_2=\Theta_2(r,r_2)=\arccos[(r^2 + r_2^2 - \eta^2)/(2r_2r)]\;\;\;
(0\le\tilde{\Theta}_2\le\pi),\eeq
whereas the azimuthal angle $0\le\phi_2\le 2\pi$. The distance $r_2$ varies in the range
\beq r_{2{\mbox{\tiny min}}}\le r_2\le r+\eta,\eeq
where $r_{2{\mbox{\tiny min}}}=\max(R,r-\eta)$. Thus the
integration with respect to ${\bf r_2}$ in eqs.(8) reduces to integration with respect to 
$\varphi_2$ and $r_2$ with fixed $\Theta_2=\tilde{\Theta}_2$.

For given locations of beads 1 and 2, the possible locations of beads 3 and 4  lie on circles of
radius $r_0=\eta\sin\beta_0$ with their location and orientation completely determined by the
coordinates of beads 1 and 2. This is due to the constraints that all bond angles are equal to
$\beta_0$ and all bond lengths are equal to $\eta$. Due to the same constraints, if the locations
of beads 2 and 4 are given, the possible locations of bead 6 lie on a circle  of radius $r_0$,
whereof the location and orientation are completely determined by the coordinates of beads 2 and 4.
Further, for  the given locations of beads 1 and 3, the possible locations of bead 5 lie on a 
circle of radius $r_0$ with the position and orientation 
completely determined by the coordinates of beads 1 and 3.
Finally, for the  given locations of beads 3 and 5, the possible locations of bead 7 are on a 
circle of radius $r_0$,  with the location and orientation  completely determined by
the coordinates of beads 3 and 5.

Let us consider bead $s$ with unknown coordinates and two other beads, $c$ and $n$ (closest to and next
to the closest to bead $s$) with known coordinates $x_c,y_c,z_c$ and $x_n,y_n,z_n$. For example, if
$s=7$, then $c=5,n=3$;  if $s=4$, then $c=2,n=1$. Bead $s$ lies on a circle of  radius $r_0$ with the
coordinates of the center  
\beq x_0\equiv x_0(x_n, y_n, z_n, x_c, y_c, z_c)=x_n+(x_c - x_n)\frac{\eta (1+|\cos\Theta_0|)}
{\sqrt{(x_c - x_n)^2 + (y_c - y_n)^2 + (z_c - z_n)^2}}, \eeq
\beq y_0\equiv y_0(x_n, y_n, z_n, x_c, y_c, z_c)= 
y_n + (y_c - y_n)\frac{\eta (1+|\cos\Theta_0|)}
{\sqrt{(x_c - x_n)^2 + (y_c - y_n)^2 + (z_c - z_n)^2}}, \eeq
\beq z_0\equiv z_0(x_n, y_n, z_n, x_c, y_c, z_c)= 
z_n + (z_c - z_n)\frac{\eta (1+|\cos\Theta_0|)}
{\sqrt{(x_c - x_n)^2 + (y_c - y_n)^2 + (z_c - z_n)^2}}.
\eeq
The coordinate $x_s$ of bead $s$ can change in the range 
\beq x_c-x_b\le x_s\le x_c+x_b\;\;\;\;(s=3,...,7),\eeq 
where 
\beq x_b\equiv x_b(x_n, y_n, z_n, x_c, y_c, z_c)= 
\eta\sin{\Theta_0}\sqrt{\frac{((y_c - y_0)^2+(z_c-z_0)^2)}{(x_c - x_n)^2 + (y_c - y_n)^2 + 
(z_c - z_n)^2}}.\eeq

For a given $x_s$, the coordinate $y_s$ of bead $s$ can have only one of two values, 
\be y^{\pm}_s&\equiv& y^{\pm}_s(x_s, x_n, y_n, z_n, x_c, y_c, z_c) = y_0+ 
\frac{(-(x_c - x_0)(y_c - y_0)(x_s-x_0)}{(y_c - y_0)^2+(z_c-z_0)^2}\pm \\
&&\frac{|z_c - z_0| 
\sqrt{[(y_c - y_0)^2+(z_c-z_0)^2]\eta^2\sin^2\beta_0-[(x_c - x_n)^2 + (y_c - y_n)^2 + (z_c - z_n)^2]
(x_s - x_0)^2}}{(y_c - y_0)^2+(z_c-z_0)^2},\nonumber\ee
For given $x_s$ and $y_s$, the coordinate $z_s$ 
of bead $s$ can have only a single value 
\be z_s^{\pm}\equiv z_s(x_s, y_s^{\pm}, x_n, y_n, z_n, x_c, y_c, z_c)=
z_0-\frac{(x_c - x_0)(x_s-x_0)+(y_c - y_0)(y_s-y_0)}{z_c - z_0}. 
\ee 

Thus, the probability distribution function $p({\bf r},{\bf r_2},{\bf r_3},{\bf r_4},{\bf r_5},{\bf r_6},{\bf r_7})$ acquires the form 
\be p({\bf r},{\bf r_2},{\bf r_3},{\bf r_4},{\bf r_5},{\bf r_6},{\bf r_7})
&=&f^{-1}\delta (\Theta_2-\tilde{\Theta}_2)\Pi_{i=3}^{7}[\delta (y_i-y_i^{-})+\delta (y_i-y_i^{+})]
\delta (z_i-\tilde{z}_i)\times\nonumber\\
&&\exp[-\phi'_{\delta}(r,({\bf r},{\bf r_2},{\bf r_3},{\bf r_4},{\bf r_5},
{\bf r_6},{\bf r_7})],\ee
where $f$ is a normalization constant determined by the condition 
\beq 
\int_{\Omega_{18}}d{\bf r_2}d{\bf r_3}d{\bf r_4}d{\bf r_5}d{\bf r_6}d{\bf r_7}\; 
p({\bf r},{\bf r_2},{\bf r_3},{\bf r_4},{\bf r_5},{\bf r_6},{\bf r_7})=1.
\eeq 
Substituting eq.(19) into eq.(8) reduces an 18-fold integral to a 7-fold one: 
\be
\bar{\phi}_{\delta}(r)&=&f^{-1}\sum_{i,j,k,m,n=+,-}\int_{0}^{2\pi}d\varphi_2\int_{L_2}dr_2\int_{L^i_3}dx_3
\int_{L^j_4}dx_4
\int_{L^k_5}dx_5\int_{L^m_6}dx_6\int_{L^n_7}dx_7\times\nonumber\\ 
&&r_2^2\,\sin{\Theta_2(r,r_2)}\,\tilde{\phi}_{ijkmn}(\varphi_2,r_2,x_3,...,x_7)
\exp[-\tilde{\phi}_{ijkmn}(\varphi_2,r_2,x_3,...,x_7)/k_BT],
\ee 
\be
f&=&\sum_{i,j,k,m,n=+,-}\int_{0}^{2\pi}d\varphi_2\int_{L_2}dr_2\int_{L^i_3}dx_3
\int_{L^j_4}dx_4
\int_{L^k_5}dx_5\int_{L^m_6}dx_6\int_{L^n_7}dx_7\times\nonumber\\ 
&&r_2^2\,\sin{\Theta_2(r,r_2)}\,\exp[-\tilde{\phi}_{ijkmn}(\varphi_2,r_2,x_3,...,x_7)/k_BT].
\ee 
In these equations each of the summation indices takes on two values, $+$ and $-$, so that there are
$5^2$ terms in the sum differing by the integrand as well as by the integration ranges (except for $L_2$
which is independent of $i,j,k,m,n$). 

The double inequalities (11) and (15) and eqs.(16)-(18) completely determine the integration ranges
(subject to the additionl constraint $x_s^2+y_s^2+z_s^2>R^2\;\;(s=2,...,7)$ of excluded cluster volume)
in  eqs.(21) and (22). In order to transform the function 
$\phi'_{\delta}(r)\equiv\phi'_{\delta}(r,{\bf
r_2},{\bf r_3},{\bf r_4},{\bf r_5},{\bf r_6},{\bf r_7})$ into the  function
$\tilde{\phi}_{ijkmn}(r,\varphi_2,r_2,x_3,...,x_7)$, the variables $y_s\;\;(s=3,...,7)$ and
$z_s\;\;(s=3,...,7)$ in the former must be replaced by $y^{\pm}_s$ and  $z^{\pm}_s$ in all the possible
combinations each of which gives rise to a  term in the sums in eqs.(21) and (22) (note that
$z^{+}_s=z_s(x_s,y_s^+,...)$ and $z^{-}_s=z_s(x_s,y_s^-,...)$). The polar  angle of bead $2$ must be
replaced by $\tilde{\Theta_2}=\Theta_2(r,r_2)$.

For example, consider the term with $i=+,j=-,k=+,m=+,n=-$ in the sum in eqs.(21),(22). The 
corresponding integrand  $\tilde{\phi}_{+-++-}(\varphi_2,r_2,x_3,...,x_7)$ is obtained  from \\ 
$\phi_{\delta}(r,{\bf r_2},{\bf r_3},{\bf r_4},{\bf r_5},{\bf r_6},{\bf r_7})$ as follows:\\ 
$y_3$ must be replaced by $y_3^+\equiv y^+_3(x_3,x_2,y_2,z_2,x_1,y_1,z_1)$ and 
$z_3$ by $z_3^+\equiv z_3(x_3,y_3^{+},x_2,y_2,z_2,x_1,y_1,z_1)$,
$y_4$ must be replaced by $y_4^-\equiv y^-_4(x_4,,x_1,y_1,z_1,x_2,y_2,z_2)$ and 
$z_4$ by $z_4^-\equiv z_4(x_4,y_4^{-},x_1,y_1,z_1,x_2,y_2,z_2)$,
$y_5$ must be replaced by $y_3^+\equiv y^+_5(x_5,x_1,y_1,z_1,x_3,y_3,z_3)$ and 
$z_5$ by $z_5^+\equiv z_5(x_5,y_5^{+},x_1,y_1,z_1,x_3,y_3,z_3)$,
$y_6$ must be replaced by $y_3^+\equiv y^+_6(x_6,x_2,y_2,z_2,x_4,y_4,z_4)$ and 
$z_6$ by $z_6^+\equiv z_6(x_6,y_6^{+},x_2,y_2,z_2,x_4,y_4,z_4)$,
$y_7$ must be replaced by $y_7^-\equiv y^-_7(x_7,x_3,y_3,z_3,x_5,y_5,z_5)$ and 
$z_7$ by $z_7^-\equiv z_7(x_7,y_7^{-},x_3,y_3,z_3,x_5,y_5,z_5)$,

Figure 2 presents $\phi_{\delta}(r)$, the contribution from the average dihedral potential to the total
potential around the cluster, provided by eqs.(21),(22). It  has quite a remarkable behavior. Starting
with its maximum value at the cluster surface, it  monotonically decreases with increasing $r$ until it
becomes constant for some  $r\ge \tilde{r}$ (see section 4). This behaviour can be accounted for by the
entropic effect on the average dihedral potential assigned to a selected bead. Actually, the closer the
selected bead ($1$) is to the cluster surface (for $r<\tilde{r}$), the more restricted is the
configurational space available for the neighboring beads (2 through 7).  This decreases the entropy of
the heteropolymer chain compared to the case where bead is far enough away from the  cluster  which, in
turn transpires as an increase in the average dihedral potential $\phi_{\delta}(r)$, assigned to
bead 1, with decreasing $r$.  With some degree of liberty, $\phi_{\delta}(r)$ can be interpreted
as a constrained (with bead 1 fixed)  free energy of the heteropolymer chain $1$ through $7$.

\subsection{Determination of the emission and absorption rates}

Figure 3 presents typical shapes of the constituents $\phi_i(r)$ and  $\phi_{\delta}(r)$ of the
potential well as functions of the distance from the cluster center,  as well as the overall potential
well $\psi_i(r)$ itself (for details of numerical calculations see section 4).   The contribution
$\phi_i(r)$, arising from the pairwise interactions, has a familiar form$^{20-25}$ reminiscent of the
underlying Lennard-Jones potential. Its combination with the  contribution $\phi_{\delta}(r)$ from the
average dihedral potential  results in the overall potential $\psi_i(r)$ which has a double well shape:
the inner well is separated by  the potential barrier from the outer well.  This shape of  $\psi_i(r)$
is of crucial importance to our model for the nucleation mechanism of protein folding  because  it makes
it possible to use a mean first passage time analysis for the determination of the rate of  absorption
of beads by the cluster 

Developing our model in the spirit of the mean first passage time analysis,$^{20-25}$ 
a bead (residue) is considered as belonging to the cluster as long as it remains in the inner 
potential well (hereinafter referred to as ``i.p.w."), 
and as  dissociated from the cluster w hen it passes over the barrier between the i.p.w. and the outer
potential well (hereinafter referred to as ``o.p.w."). 
The rate of emission, $W^-$, is determined by the mean time necessary for the passage of the bead 
from the i.p.w. over the barrier into the o.p.w. Likewise, a bead is considered as belonging to the unfolded
part of the heteropolymer (protein) as long as it remains in the o.p.w., 
and as absorbed by the cluster when it passes over the barrier between the o.p.w. and the i.p.w.. 
The rate of emission, $W^+$, is determined by the mean time necessary for the passage of the bead 
from the o.p.w. over the barrier into the i.p.w. 

\par The mean first passage time of a bead escaping from some potential well is calculated on the basis of  a
kinetic equation governing the chaotic motion of the bead in that potential well. 
The chaotic motion of the bead  is assumed to be 
governed by the Fokker-Planck equation for the single-particle distribution function with respect to
its coordinates and momenta, i.e., in the phase space.$^{47-49}$ 
Prior to the passage event, the 
evolution of a bead in both the i.p.w. and o.p.w. occurs in a dense enough medium 
(cluster folded residues or unfolded but compact part of the protein), where 
the relaxation time for its velocity  distribution function is extremely short and negligible compared
to the characteristic time  scale of the passage process. 
Under favorable conditions, the 
Fokker-Planck equation reduces  to the Smoluchowski equation, which involves diffusion in an external
field.$^{48,49}$ Solving that equation,one can obtain$^{24}$ the following expressions for $W^-$ and $W^+$, 
the emission and
absorption rates, respectively: 
\beq W^-=n_{\mbox{\small iw}}D_{\mbox{\small iw}} 
\omega_{\mbox{\small iw}},\;\;\;\;\; 
W^+=n_{\mbox{\small ow}}D_{\mbox{\small ow}} \omega_{\mbox{\small ow}}.\eeq  
Here the subscripts ``iw" and ``ow" mark the quantities for the inner and outer potential wells,
respectively; $n$ is the number of beads in the well, $D$ is the diffusion coefficient of the bead, and 
$\omega_{\mbox{\small iw}}$ and $\omega_{\mbox{\small ow}}$ are defined as 
\beq \omega_{\mbox{\small iw}}\equiv 1/D_{\mbox{\small iw}}\bar{\tau}_{\mbox{\small iw}},\;\;\;
\omega_{\mbox{\small ow}}\equiv 1/D_{\mbox{\small ow}}\bar{\tau}_{\mbox{\small ow}},\eeq   
where $\bar{\tau}_{\mbox{\small iw}}$ and $\bar{\tau}_{\mbox{\small iw}}$ are 
the mean first passage times for a 
bead in the i.p.w. to cross over the barrier into the o.p.w. and  vice-versa, respectively. 
Note that in the original work$^{20-22}$ on the mean first passage time analysis in the nucleation theory
and its recent development$^{23-25}$ this method was used only for determining $W^-$ but not $W^+$. 
In the present model, the 
double well character of the overall potential $\psi(r)$ around the cluster allows one to apply the 
first passage time analysis also to determining $W^+$.

Clearly, the quantities $W^-,W^+,\omega_{\mbox{\small iw}}, \omega_{\mbox{\small iw}},
\bar{\tau}_{\mbox{\small iw}},\bar{\tau}_{\mbox{\small iw}}$ are the functions of the cluster size and
composition (for the explicit form see ref.24). However, since the overall composition of the
protein (heteropolymer)  is fixed, one can assume that the cluster which forms during its folding has a
constant composition equal to the overall protein composition which leads to a unary nucleation
theory.  (Strictly speaking, one can develop a theoretical model for the nucleation mechanism of
protein folding without this assumption which would lead to a binary nucleation theory, but this would
drastically complicate the problem computationally.) Under this assumptions the aforementioned
quantities are functions of only the size of the cluster (say, its radius $R$ or the total number of
beads $\nu$ therein).

\subsection{The equilibrium distribution and steady-state nucleation rate}

\par In terms of nucleation, during the protein folding clusters of various sizes may emerge and exist
simultaneously with different probabilities. Let us denote the distribution of clusters with respect to
the number of beads in a cluster at time $t$ by $g(\nu,t)$.  Once the emission and absorption rates
$W^-=W^-(\nu)$ and $W^+=W^+(\nu)$ are known as functions of the cluster size, one can find the
equilibrium distribution of clusters  $g_e(\nu,t)$  and solve the kinetic equation of nucleation to find
the steady-state nucleation rate.  

Actually, according to the principle of detailed balance, 
$ W^+(\nu-1)g_e(\nu-1)=W^-(\nu)g_e(\nu)$,  
which can be rewritten as 
\beq \frac{g_e(\nu)}{g_e(\nu-1)}=\frac{W^+(\nu-1)}{W^-(\nu)}.\eeq
By applying eq.(25) to $(\nu-i)$ with $i=2,3,...,\nu-1$, 
multiplying the RHSs and LHSs of all equalities, one obtains 
\beq \frac{g_e(\nu)}{g_e(1)}=
\prod_{i=1}^{\nu-1}\frac{W^+(\nu-i)}{W^-(\nu-i+1)}.\eeq
The equilibrium distribution of clusters $\nu=1$ is just the number density of residues in a compact
(but unfolded) protein, i.e.,$g_e(1)=\rho_u$, so that  equation (18) can be rewritten as 
\beq g_e(\nu)=\rho_u \frac{W^+(1)}{W^+(\nu)}\prod_{i=1}^{\nu-1}\frac{W^+(\nu-i+1)}{W^-(\nu-i+1)}.\eeq  

Let us introduce the function $G(\nu)=-k_BT\ln[{g_e(\nu)/\rho_u}]$.  Clearly, $G(\nu)$ in the present
theory plays a role similar to the the free energy of cluster formation in CNT$^{50-52}$ 
Under favorable conditions, $G(\nu)$ first increases with increasing $\nu$, attains its maximum at some
$\nu=\nu_c$, and then decreases. In CNT, $G(\nu)$ has also a minimum following the maximum but only for an
NVT ensemble where the growth of the cluster (i.e., increase of $\nu$) leads to the decrease in the
metastability of the mother phase. Since the folding protein cannot be considered as an NVT ensemble, we
will not consider this case. 
The essence of our model, as an alternative to the CA-based theory,  consists of constructing the
equilibrium distribution of clusters, $g_e(\nu)$ (and the function $G(\nu)$  without employing  the
classical  thermodynamics.  

\par  The kinetic equation of nucleation in the vicinity of the critical point can be written as$^{50-52}$ 
\beq \frac{\partial g(\nu,t)}{\partial t}=W_{c}^+\frac{\partial }{\partial \nu} 
\left[\frac{\partial }{\partial \nu }+\frac{\partial G}{\partial \nu}\right]g(\nu,t).\eeq 
(subscript ``c" marks quantities at the critical point) 
and the function $G(\nu)$ can be accurately represented by its bilinear form. 
The steady-state solution of the kinetic equation (20) in the vicinity of $\nu_c$ subject to the
conventional boundary conditions  
\beq \frac{g(\nu,t)}{g_{e}(\nu)}\rightarrow
1\;\;\;\;\;(\nu\rightarrow 0),\;\;\;\;\; \frac{g(\nu,t)}{g_{e}(\nu)}\rightarrow
0\;\;\;\;\;(\nu\rightarrow \infty),\eeq  
where $g_{e}(\nu)$ is the equilibrium distribution,
provides the steady-state  nucleation rate$^{50-52}$ which can be presented in the form 
\beq J_s=\frac{W_{c}^+}{\sqrt{\pi}\De\nu_c}\rho_u\mbox{e}^{-G_c/k_BT}, \eeq  
where 
\beq\De\nu_c=\left|\frac{\partial^2 G}{\partial \nu^2}\right|_c^{-1/2}.\eeq

\subsection{Evaluation of the protein folding time}

Knowing the emission and absorption rates as functions of $\nu$ as well as the nucleation rate $J_s$,
one can estimate the time $t_f$ necessary for the protein to fold via nucleation. To do so we will
regard the protein folding (via nucleation) as a two stage process. At the first stage, a critical
cluster of native residues form(nucleation proper). At this stage, i.e., for $\nu<\nu_c$, the emission
rate $W^-$ is larger than $W^+$, but the cluster does attain the critical size by means of
fluctuations.  At the second stage the nucleus grows via regular absorption of native residues
dominating their emission, $W^-<W^+$ for $\nu>\nu_c$. Thus, the folding time can be represented as 
\beq t_f\simeq t_n+t_g,\eeq
where $t_n$ is the time necessary for one critical cluster to nucleate within a compact (but still
unfolded) protein and $t_g$ is time necessary for the nucleus to grow up to the maximum size, i.e.,
attain the size of a folded protein. 

The time $t_n$ of the first nucleation event can be estimated  as
\beq t_n\simeq 1/J_sV_0,\eeq
where $V_0$ is the volume of the unfolded protein in a compact state. The growth time $t_g$ can be found by
solving the differential equation
\beq \frac{d\nu}{d t}=W^+(\nu)-W^-(\nu)\eeq
subject to the initial condition $\nu=\nu_c$ at $t=0$ and the condition $\nu=N_0$ at $t=t_n$. The solution
of eq.(34) is given by the integral
\beq t_n\simeq \int_{\nu_c}^{N_0}\frac{d\nu}{W^+(\nu)-W^-(\nu)}.\eeq

\section{Numerical evaluations}

In this section we will present some numerical results of the  application of our model to the folding
of a model protein, namely, a heteropolymer consisting of total 2500  hydrophobic and hydrophilic
residues, with the mole fraction of hydrophobic residues $\chi_0=0.75$. The interactions between a pair
of non-linked beads  were modeled via the Lennard-Jones (LJ) type potentials (1), while the potential
due to the dihedral angle $\delta$ was modeled according to eq.(3). The presence of water molecules was
not taken into account explicitly but was assumed to be implemented into the model via the potential
parameter. 

\par All numerical calculations were carried out for the following values of the interaction 
parameters: 
$$ \eta_{1}=5.39\times 10^{-8} \mbox{cm},\;\;\;
\epsilon_{l}=(2/700)\epsilon_{b},\;\epsilon'_{\delta}=
\epsilon''_{\delta}=0.3\epsilon_{b},\;\epsilon_{b}/kT=1.$$ 
A typical
density of the the folded protein was evaluated according to data in refs.53,54 and was set to 
$\rho_f\eta^3=1.05$, while a typical density of the unfolded protein in the compact configuration was
set to be $\rho_u=0.25\rho_f$ (note that similar values for $\rho_f$ and $\rho_d$ are suggested  in
ref.18). Taking into account the results in ref.55,  the diffusion coefficients in the i.p.w. and the
o.p.w.  were assumed to be related as $D_{\mbox{\tiny iw}}\rho_f=D_{\mbox{\tiny ow}}\rho_u$. Because
of the lack of reliable data on the diffusion coefficient of a residue in a protein chain, 
$D_{\mbox{\tiny iw}}$ was assumed to vary between $10^{-6}$ cm$^2$/s and $10^{-8}$ cm$^2$/s.

\par Figure 2 shows the average dihedral potential (assigned to a selected bead)
$\bar{\phi}^{\delta}(r)$  as a function of $r$ for three clusters of sizes (a) $R=3\eta$, (b)
$R=6\eta$, and (c) $R=9\eta$. The points represent the actual numerical results obtained  by using $1\times
10^6$ to $2\times 10^6$ point Monte Carlo integration in calculating 7-fold integrals in eqs.(11),(12).
The vertical dashed lines correspond to $r'$ (see above) such that  $\bar{\phi}^{\delta}(r)$ is
expected to be constant for $r>r'$. The solid lines are analytic fits by an expression
$a+b\exp[-c(r-d)^2]$. With the accuracy of our calculations the parameters $a,b,$ and $c$ of this fit
do not change with $R$, while the parameter $d$ is roughly $R+\eta$. Clearly, with an increased accuracy
of calculations we may eventually find some dependence of $a,b,c$ on $R$, but with our current accuracy
it appears that  $\bar{\phi}^{\delta}(r)$ has an universal shape independent on $R$. Undoubtedly, the
calculation of  $\bar{\phi}^{\delta}(r)$ will constitute  the most time consuming procedure in applying
our model to real life problems. It takes about 24 hours to  obtain one value of 
$\bar{\phi}^{\delta}(r)$ (one point in Fig.2)  on a Dell/Pentium4/3Ghz/512Mb computer. 

\par Figure 3 presents typical shapes of the potentials $\phi_b(r)$ (lower solid curve),  
$\bar{\phi}^{\delta}(r)$ (upper solid curve), and $\psi_b(r)$ (dashed curve) for a hydrophobic bead 
around the cluster  as functions of the distance $r$  from the center of the cluster of radius
$R=3\eta$.  The potential $\phi_i(r)$ is due to the pairwise interactions of the Lennard-Jones type, and
has a shape reminiscent thereof. Previous applications$^{20-25}$  of the mean first passage time
analysis to nucleation  had invariably lead to this kind of the potential well around the cluster.  

The average dihedral potential (assigned to a selected bead)
$\bar{\phi}^{\delta}(r)$   has a maximum value at the cluster surface and decreases monotonically  with
increasing $r$ until it becomes constant for $r\ge \tilde{r}$ 
which is the maximum distance between beads 1 and
6 (or beads 1 and 7) dependent on $R,\eta,$ and $\Theta_0$:  
$$ \tilde{r}=R + \eta \left(1 + \sqrt{3-2\cos{\Theta_0}+2\sqrt{2(1 - 
\cos{\Theta_0})}\sin{\frac{\Theta_0}{2}}}\right).$$
Such a behavior of $\bar{\phi}^{\delta}(r)$  can be thought of as a consequence of an increase in the 
entropy of the heteropolymer chain as the selected bead 1 approaches the cluster  surface for $r<r'$
which occurs because the configurational space available for the neighboring beads  (2 through 7)
becomes more and more restricted. Once $r$ becomes greater than $r'$, this piece of the heteropolymer
chain (beads 1 through 7) does not feel the presence of the cluster any more. With some degree of
liberty $\bar{\phi}_{\delta}(r)$ can be interpreted as a constrained (with bead 1 fixed) 
free energy of that piece of the heteropolymer which includes beads $1$ through $7$.

As a result of the combination of $\phi_b(r)$ and $\bar{\phi}_{\delta}(r)$  the overall potential
$\psi_b(r)$ has a double well shape: the inner well is separated by the potential barrier from the
outer well. The geometric characteristics of the wells (widths, depths, etc...) and the height and
location of the barrier between them are determined by the interaction parameters
$\epsilon_{b},\epsilon_{l},\epsilon_{\delta}$. For example, the larger the ratio
$\epsilon_{\delta}/\epsilon_b$, the higher the barrier between the well, the wider the i.p.w. and the
narrower the o.p.w. Note that the barrier has different heights for beads in the i.p.w. and o.p.w. 
The outer boundary of the o.p.w. is due to the confining potential arising because all residues around
the cluster are successively linked and are bound thereto. Hence they are confined within some 
volume wherein the protein is encompassed and the location $r_{\mbox{\tiny cf}}$ of the confining
potential is assumed to coincide with its outer boundary. For a given $N_0$ the location 
$r_{\mbox{\tiny cf}}$  is determined by the
the size of the cluster and densities $\rho_f$ and $\rho_u$. The existence of the o.p.w. allows one to
consider the absorption of a bead by the cluster as an escape of the bead from the o.p.w. by crossing
over the barrier into the i.p.w. This makes it possible to use the mean first passage time analysis for
the determination of the rate of  absorption of beads by the cluster. Since the use of the traditional
expression for the absorption rate (based on the gas-kinetic theory)$^{46}$ is rather inadequate in the
cluster growth within the protein,  the double-well shape of $\psi_i(r)$ is of crucial  importance to
our model for the nucleation mechanism of protein folding.

\par Figure 4 presents $W^-$ and $W^+$, the emission and absorption rates, respectively, as functions of
the cluster size $R$. The location of the intersection of these functions determines the size of the
critical cluster, $R_c$. The emission rate is greater than the absorption rate, $W^-(r)>W^+(r)$, for
small clusters with 
$R<R_c$, whereas for clusters larger than the nucleus the absorption dominates over the emission, 
$W^-(r)<W^+(r)$ for $R>R_c$. Note that both $W^-$ and $W^+$ increase with increasing $R$, but $W^-$
increases roughly linearly with $R$ whereas $W^+$ shoots up by several orders of magnitude after the
cluster becomes supercritical. This is a consequence of the fact that the width of the o.p.w. quickly
decreases as the cluster grows while the outer height of the barrier between the i.p.w. and o.p.w. does not
not virtually change, and it becomes increasingly easy for a bead which is in the o.p.w. 
to cross over the barrier and fall into the i.p.w. 

The behaviour of $W^-$ and $W^+$ also explains our numerical estimates for the characteristic times of
the first nucleation event $t_n$, growth time $t_g$, and total folding time $t_f$ by eqs.(32),(33), and
(35). Although these depend very much on the location of $R_c$ and the value of $D_{\mbox{\tiny iw}}$ ($R_c$
itself does not depend on $D_{\mbox{\tiny iw}}$ but only on the ratio 
$D_{\mbox{\tiny ow}}/D_{\mbox{\tiny iw}}$), always 
$t_n\gg t_g$, i.e., the protein folding time is mainly determined by the time necessary for the first
nucleation event. Physically, this is the case because the increase of the cluster size from $\nu=1$ to
$\nu=\nu_c$ occurs only owing to fluctuations which have to overcome the natural tendency of a small 
cluster to decay ($W^+<W^-$ for $\nu<\nu_c$). For supercritical clusters $W^+$ so quickly immensely
overwhelms $W^-$ that fluctuations are unable to impede the natural tendency of the cluster to grow
(strictly speaking, this is true only for $\nu>\nu_c+\De\nu_c$, but for rough estimates eq.(35) is
acceptable). 
For the above choice of system parameter and $D_{\mbox{\tiny iw}}$ in the range from $10^{-6}$ cm$^2$/s 
to $10^{-8}$ cm$^2$/s our model predicts the characteristic time of the protein folding (for $N_0=2500$)
in the range from several seconds to several hundreds of seconds which is in a very good agreement with
expectations based on experimental data. 

\section{Conclusions} 

So far most of the work on the protein folding has been done by using either Monte Carlo (MC) or
molecular dynamics (MD) simulations. The rigorous theoretical  treatment of the protein folding by
means of the statistical mechanics is hardly practicable because of the extreme complexity of the
system. A number of simulations have suggested that there can exist multiple pathways for a  protein to
fold one of which has been identified as reminiscent of nucleation. However, a theoretical model for
the  nucleation mechanism of the process had so far remained underdeveloped.   The previous model,
based on the approach of the classical nucleation theory (CNT), was a purely thermodynamic one
considered the formation of a cluster of protein residues and  calculated the free  energy change
thereupon.$^{14,18}$ The number of a critical cluster (nucleus) was provided by  the location of the
maximum of  the free energy of formation as a function of a single independent variable of state of the
cluster. In such a model the free energy of cluster formation depends on the surface tension of  a
cluster of protein residues. This quantity is an ill-defined physical quantity and can be considerer
only as  an adjustable parameter. According to the nucleation mechanism, after the formation of the
nucleus (critical size cluster of residues),  the protein quickly reaches its native state. 

In the present work we present a new, microscopic model for the nucleation mechanism of the protein
folding. A protein is considered as a heteropolymer consisting of two type of beads (hydrophobic and
hydrophilic) linked with bonds of fixed length. All bond angles are also assumed to be fixed and equal
to 105$^o$. All non-adjacent beads are assumed to interact via Lennard-Jones like potential. Besides
these interactions, the total energy of the heteropolymer contains a contribution from dihedral angles 
of all triads of successive links. Unlike the old model, ours is developed without recurring to CNT
approach. Instead, it is based on the above ``molecular" interactions, both long-range and
configurational. The parameters of these potentials can be rigorously defined, unlike  the ill-defined
surface tension of a cluster of protein residues. 

The crucial idea underlying the new model consists of averaging the dihedral potential in which a
selected residues is involved over all the possible  configurations of neighboring residues. The
resulting average dihedral potential depends on the distance between the residues and the cluster
center. It has a maximum at the cluster surface and monotonically decreases with increasing distance
therefrom. Its combination with the average   potential due to pairwise interactions between the
selected residue and those in the cluster a double  potential well around the cluster with a barrier
between the two wells. Residues in the inner well are considered to belong to the cluster (part of the
protein with correct tertiary contacts)  while those in the outer well are treated as belonging to the
mother phase (amorphous  part of the protein with incorrect tertiary contacts). Transitions of residues
from the inner well into the outer one and vice versa are considered as elementary emission and
absorption events, respectively. The rates of these processes are determined by using the mean first
passage time analysis.  Once  these rates are found as functions of the cluster size, one can develop a
self-consistent kinetic  theory for the nucleation mechanism of protein folding. For example,  the size
of the critical cluster (nucleus) is then found as  the one for which these rates are equal. The time
necessary for the protein to fold can be evaluated as a sum of the times necessary for the appearance
of the first nucleus and the time necessary for the nucleus to grow to the maximum size (of the folded
protein in the native state). 

For numerical illustration we have considered a model protein consisting of $2500$ beads with the mole
fraction of hydrophobic beads equal to $0.75$. The composition of the cluster during its formation and
growth was assumed to be constant and equal to the composition of the whole protein. This allows one to
consider the model as a single component one. The size of the critical cluster and the folding time
predicted by the model depend very much on many parameters of the system, such as interaction
parameters, densities of the protein  in the unfolded (but compact) and folded states, diffusion
coefficients therein, etc. With an appropriate choice of interaction parameters and densities, the size
of the critical cluster predicted by our model is about $220$ residues and with the free energy of
nucleus formation being about $20k_BT$. This results suggest that the quantity equivalent to the
``surface tension" in the old model of nucleation in a protein, should be smaller than the previous
estimates$^{14,18}$ of the latter by an order of magnitude.  The characteristic time of protein folding
was estimated to be in the range from several seconds to several hundreds of seconds depending on the
diffusion coefficient of native residues in the range $10^{-6}$ cm$^2$/s to $10^{-8}$ cm$^2$/s. This is
consistent with experimental data$^{55}$ on typical folding times of proteins as well as  with
estimates obtained by  other theoretical models$^{18}$ and in simulations.$^{6,14}$

A further development of our model will require the removal of several simplifying assumptions that we
recurred to in the present work. For example, it would be more appropriate to model a protein as a
three-component heteropolymer (including not only hydrophobic and hydrophilic residues, but also
neutral ones). This will result in more lengthy numerical calculations of the average dihedral
potential because the dihedral potential involving neutral beads is expected to be lower and requires
separate calculations. Next, the cluster composition during its formation and growth can quite
significantly depend on the cluster size, particularly in the vicinity of the critical size, so
assuming it constant in the present work might have lead to serious inaccuracy in the results.
Including neutral beads in the model and allowing the cluster composition to differ from that of the
protein will result in a binary or even ternary nucleation mechanism of protein folding. However,
besides some increase in computational efforts there seem to exist no principal difficulty in
developing the model in these directions. Some other improvements of the model can be also introduced
in the model, but they will be discussed in our future papers on the subject.

{\em Acknowledgments} - {\small I am grateful to Professors F.M.Kuni, A.P.Grinin, and E.Ruckenstein for
many helpful discussions of this work which was 
was supported by the National Science Foundation through  the
grant CTS-0000548}.

\newpage
\section*{References}
\begin{list}{}{\labelwidth 0cm \itemindent-\leftmargin} 
\item $^{1}$T.E.Creighton, in {\it Proteins: Structure and Molecular Properties}, edited by 
	W.H.Freeman (San Francisco, 1984) 
\item $^{2}$L.Stryer, in {\it Biochemistry}, edited by W.H.Freeman  (1988).
\item $^{3}$C.Ghelis and J.Yan, {\it Protein Folding} (Academic Press, New York,1982).
\item $^{4}$C.B.Anfinsen, Science {\bf 181}, 223-230 (1973).
\item $^{5}$J.D.Honeycutt and D.Thirumalai, Proc.Natl.Acad.Sci. USA {\bf 87}, 3526-3529 (1990).
\item $^{6}$J.D.Honeycutt and D.Thirumalai, Biopolymers {\bf 32}, 695-709 (1992).
\item $^{7}$J.S.Weissman and P.S.Kim, Science {\bf 253}, 1386-1393 (1991) .
\item $^{8}$T.E.Creighton, Nature (London) {\bf 356}, 194-195 (1992).
\item $^{9}$P.S.Kim and R.I.Baldwin, Annu.Rev.Biochem. {\bf 51}, 459 (1982).
\item $^{10}$P.S.Kim and R.I.Baldwin, Annu.Rev.Biochem. {\bf 59}, 631 (1990).
\item $^{11}$T.E.Creighton, Biochem. J. {\bf 240}, 1 (1990).
\item $^{12}$T.E.Creighton, Prog.Biophys.Mol.Biol. {\bf 33}, 231 (1978). 
\item $^{13}$Z.Guo, D.Thirumalai, and J.D.Honeycutt, J.Chem.Phys. {\bf 97}, 525-535 (1992).
\item $^{14}$Z.Guo and D.Thirumalai, Biopolymers {\bf 36}, 83-102 (1995).
\item $^{15}$D.Thirumalai and Z. Guo, Biopolymers {\bf 35}, 137-140 (1995).
\item $^{16.}$J.D.Bryngelson and P.G. Wolynes, Proc.Natl.Acad.Sci.(USA) {\bf 84}, 7524 (1987); 
    	J.Phys.Chem. {\bf 93}, 6902 (1998).
\item $^{17}$E.I.Shakhnovich and A.M.Gutin, Nature 346, 773 (1990); J.Phys.A {\bf 22}, 1647 (1989).
\item $^{18}$J.D.Bryngelson and P.G.Wolynes, Biopolymers {\bf 30}, 177 (1990).
\item $^{19}$K.A.Dill, Biochemistry {\bf 24}, 1501 (1985); {\bf 29}, 7133 (1990).
\item $^{20}$G.Narsimhan and E. Ruckenstein, J. Colloid Interface Sci. {\bf 128}, 549 (1989).
\item $^{21}$E.Ruckenstein and B. Nowakowski, J. Colloid Interface Sci. {\bf 137}, 583 (1990).
\item $^{22}$B.Nowakowski and E. Ruckenstein, J. Colloid Interface Sci. {\bf 139}, 500 (1990).
\item $^{23}$Y.S.Djikaev and E.Ruckenstein, J.Chem.Phys. {\bf 123}, 214503 (2005).
\item $^{24}$Y.S.Djikaev and E.Ruckenstein, J.Chem.Phys. {\bf 124}, 124521 (2006).
\item $^{25}$Y.S.Djikaev and E.Ruckenstein, J.Chem.Phys. {\bf 124}, 124521 (2006).
\item $^{26}$M.Levitt and R. Sharon, Proc.Natl.Acad.Sci.USA {\bf 85}, 8557 (1988). 
\item $^{27}$J.Skolnik, A. Kolinski, and R. Yaris, Biopolymers {\bf 28}, 1058 (1989).
\item $^{28}$A.Kolinski, J. Skolnik, and R. Yaris, Biopolymers {\bf 26}, 937 (1987).
\item $^{29}$A.Sikorski and J. Skolnik, Biopolymers {\bf 28}, 1097 (1989); 
	ibid J.Mol.Biol. {\bf 213}, 183 (1990).
\item $^{30}$J.Skolnik and A. Kolinski, Science {\bf 250}, 1121 (1990).
\item $^{31}$C.Levinthal, in {\it M\"ossbauer Spectroscopy in Biological Systems}, edited by 
	P.Debrunner, J.C.M.Tsibris and E.M\"onck, (University of Illinois Press, 1968).
\item $^{32}$D.Wetlaufer, Proc.Natl.Acad.Sci. USA {\bf 70}, 697 (1973).
\item $^{33}$T.Y.Tsong, R.Baldwin, and P.McPhie, J.Mol.Biol. {\bf 63}, 453 (1972).
\item $^{34}$J.Moult and R.Unger, Biochemistry {\bf 30}, 3816 (1991).
\item $^{35}$K.Dill, K.Fiebig, and H.S.Chan, Proc.Natl.Acad.Sci.USA {\bf 90}, 1942 (1993).
\item $^{36}$D.J.Lee, M. M. Telo da Gama, and K. E. Gubbins, J. Chem. Phys. {\bf 85}, 490 (1986).
\item $^{37}$A.I.Rusanov, {\em Phasengleichgewichte und Grenzflachenerscheinungen}
	(Academie, Berlin, 1978). 
\item $^{38}$R.C.Tolman, J. Chem. Phys., {\bf 17}, 333 (1949).
\item $^{39}$J.G.Kirkwood and F.P. Buff, J. Chem. Phys., {\bf 17}, 338 (1949).
\item $^{40}$C.Flageollet, M. Dihn Cao, and P. Mirabel, J. Chem. Phys. {\bf 72}, 544 (1980).
\item $^{41}$G.Wilemski, J. Phys. Chem. {\bf 91}, 2492 (1987).
\item $^{42}$B.E.Wyslouzil, J. H. Seinfeld, R. C. Flagan, and K. Okuyama,  
	J. Chem. Phys. {\bf 94}, 6827 (1991).
\item $^{43}$A.Laaksonen, J. Chem. Phys. {\bf 97}, 1983 (1992).
\item $^{44}$G.Wilemski, J. Chem. Phys. {\bf 80}, 1370 (1984).
\item $^{45}$Y.S.Djikaev, I. Napari, A. Laaksonen, J. Chem. Phys. {\bf 120}, 9752 (2004).
\item $^{46}$F.F.Abraham, {\it Homogeneous Nucleation Theory} (Academic, New York, 1974).
\item $^{47}$S.Chandrasekhar, Rev. Mod. Phys., {\bf 15} (1949) 1.
\item $^{48}$C.W. Gardiner, {\it Handbook of Stochastic Methods}  (Springer, New York/Berlin, 1983).
\item $^{49}$N.Agmon, J. Chem. Phys., {\bf 81}, 3644 (1984).
\item $^{50}$J.Lothe and G. M. J. Pound, In {\it Nucleation}, edited by A. C. Zettlemoyer 
	(Marcel-Dekker, New York, 1969). 
\item $^{51}$J.Schmelzer, G. R\"opke, and V. B. Priezzhev, Eds. {\it Nucleation Theory and
	Applications} (JINR, Dubna, 1999).
\item $^{52}$D.Kashchiev, {\it Nucleation: Basic Theory with Applications} 
(Butterworth-Heinemann, Oxford, 2000).
\item $^{53}$Y.Harpaz, M.Gerstein, and C.Chothia, Structure, {\bf 2}, 641-649 (1994).
\item $^{54}$D.M.Huang and D.Chandler, Proc.Natl.Acad.Sci.USA, {\bf 15}, 8324-8327 (2000).
\item $^{55}$B.N\"otling, Protein Folding Kinetics (Springer-Verlag, Berlin, 2006).  
\end{list}
\newpage
\begin{figure}[htp]
\begin{center}\vspace{1cm}
\includegraphics[width=7.9cm]{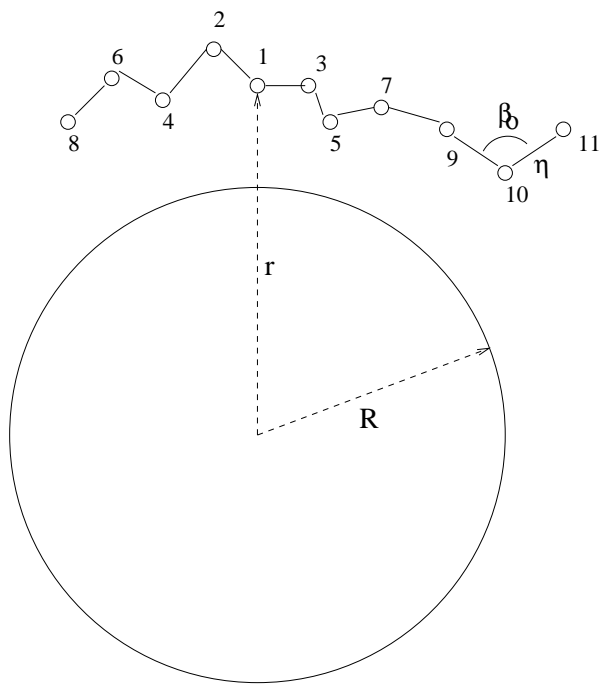}\\ [3.7cm]
\caption{\small A scheme of a piece of a heteroplymer chain around the spherical cluster (shown only partly) 
of radius $R$. Bead $1$ lies in the Fugure plane, whereas beads $2$ through $7$ may all lie in different planes, but all bond angles are equal to
$105^o$ and their lengths are equal to $\eta$.  The distance between the selected bead $1$ and the
center of the cluster is $r$}
\end{center}
\end{figure}

\newpage
\begin{figure}[htp]\vspace{-1cm}
	      \begin{center}
$
\begin{array}{c@{\hspace{0.3cm}}c} 
%	      \begin{flushright}
              \leavevmode
%      	      \hspace{-3.7cm}
      	      \vspace{1.3cm}
	\leavevmode\hbox{a) \vspace{3cm}} &   
\includegraphics[width=7.9cm]{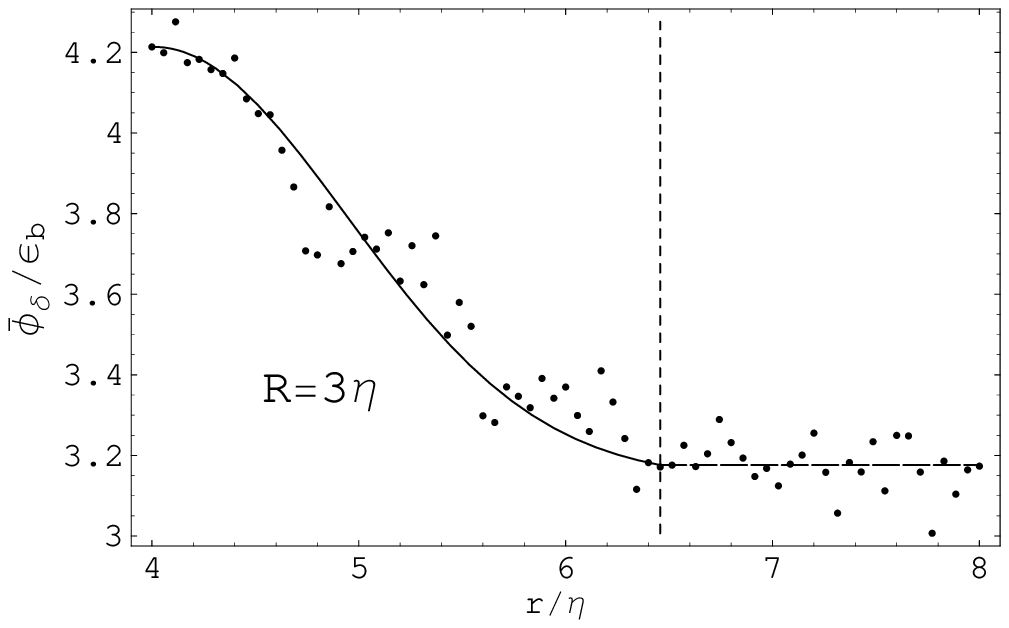}\\ [0.5cm] 
      	      \vspace{1.3cm}
	\leavevmode\hbox{b) \vspace{3cm}} &   
\includegraphics[width=7.9cm]{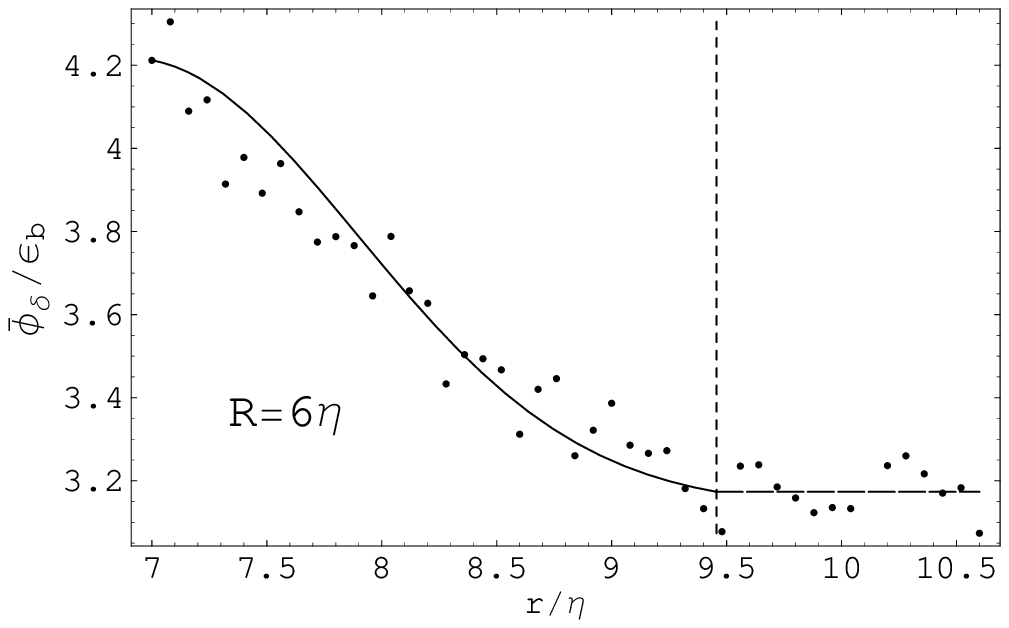}\\ [0.5cm] 
      	      \vspace{1.7cm}
	\leavevmode\hbox{c) \vspace{3cm}} &  
      	      \vspace{0.0cm}
\includegraphics[width=7.9cm]{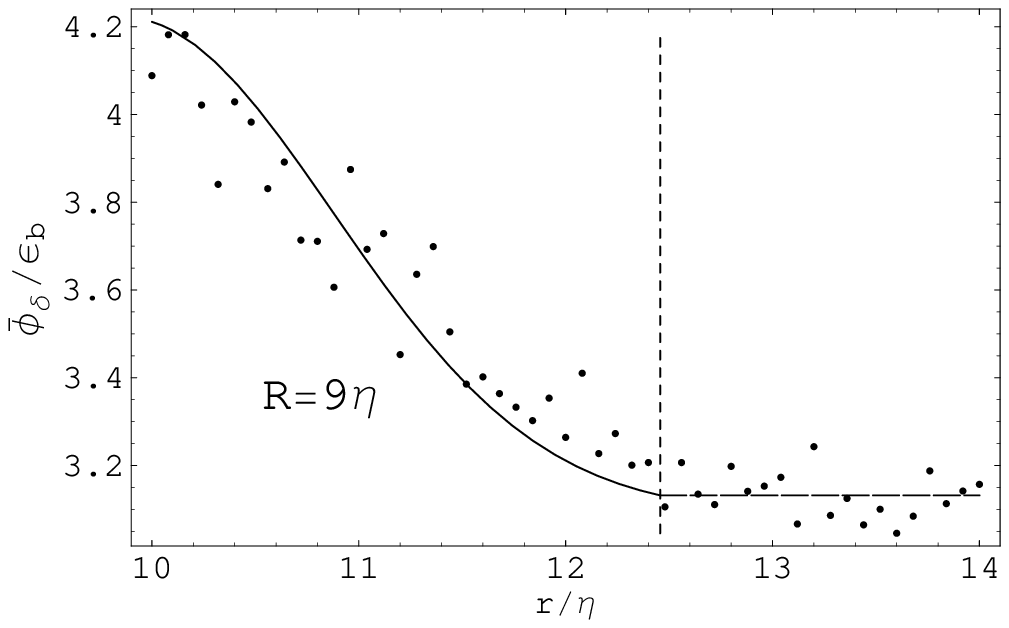}\\ [-0.7cm] 
%              \end{flushright} 
%\mbox{\bf (aa)} & \mbox{\bf (bb)} 
\end{array}  
$  

	      \end{center} 
            \caption{\small The average dihedral potential (assigned to a selected bead)
$\bar{\phi}^{\delta}(r)$  as a function of $r$ for three clusters of sizes (a) $R=3\eta$, (b)
$R=6\eta$, and (c) $R=9\eta$. The points represent the actual numerical results obtained  by using 
the Monte Carlo integration in eqs.(21),(22).
The vertical dashed lines correspond to $\tilde{r}$. The solid lines are analytic fits by an expression
$a+b\exp[-c(r-d)^2]$.} 
\end{figure}

\newpage
\begin{figure}[htp]
\begin{center}\vspace{1cm}
\includegraphics[width=7.9cm]{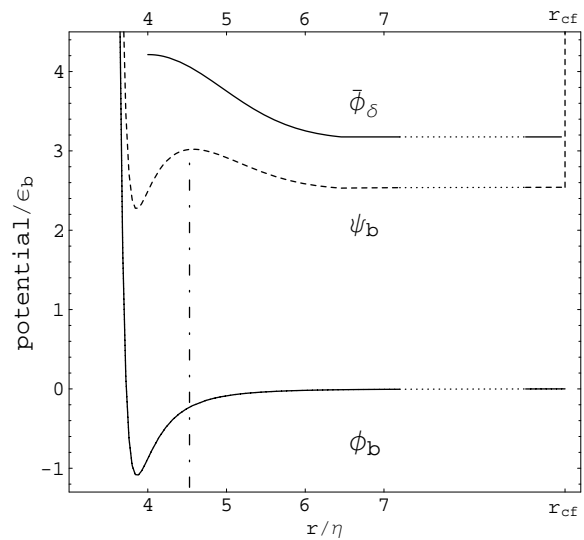}\\ [3.7cm]
\caption{\small Typical shapes of the potentials $\phi_b(r)$ (lower solid curve),  
$\bar{\phi}_{\delta}(r)$ (upper solid curve), and $\psi_b(r)$ (dashed curve) for a hydrophobic bead 
around the cluster  as functions of the distance $r$ 
from the center of the cluster of radius $R=3\eta$. The outer boundary of the
o.p.w. ($r_{\mbox{\tiny cf}}\simeq 12.99\eta$) was assumed to coincides with the outer  boundary  
of the volume wherein the whole protein is encompassed. }
\end{center}
\end{figure}
								    
\newpage
\begin{figure}[htp]
\begin{center}\vspace{1cm}
\includegraphics[width=7.9cm]{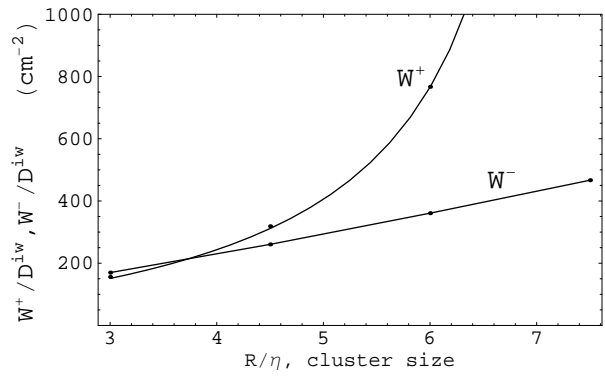}\\ [3.7cm]
\caption{\small Figure 4 presents $W^-$ and $W^+$, the emission and absorption rates, respectively, as functions of
the cluster size $R$. The location of the intersection of these functions determines the size of the
critical cluster, $R_c$}
\end{center}
\end{figure}
		
\end{document}